\documentclass[preprint,showpacs,preprintnumbers,amsmath,amssymb,showkeys]{revtex4}
 
\usepackage{graphicx}
\usepackage{dcolumn}
\usepackage{bm}
\usepackage{braket}
\usepackage{amsmath}
  
\begin{document} 

\noindent

\preprint{}

\title{Epistemically restricted phase space representation, weak momentum value, and reconstruction of quantum wave function}
 
\author{Agung Budiyono}
\email{agungbymlati@gmail.com} 

\affiliation{Research Center for the Nanoscience and Nanotechnology, Bandung Institute of Technology, Bandung 40132, Indonesia}
\affiliation{Edelstein Center, Hebrew University of Jerusalem, Jerusalem 91904 Israel}
\affiliation{Department of Engineering Physics and Research Center for Nano Science and Technology, Bandung Institute of Technology, Bandung 40132, Indonesia}
\affiliation{Kubus Computing and Research, Juwana, Pati 59185, Indonesia}
 
\date{\today}
  
\begin{abstract}   
 
A phase space distribution associated with a quantum state was previously proposed, which incorporates a specific epistemic restriction parameterized by a global random variable on the order of Planck constant, transparently manifesting quantum uncertainty in phase space. Here we show that the epistemically restricted phase space (ERPS) distribution can be determined via weak measurement of momentum followed by post-selection on position. In the ERPS representation, the phase and amplitude of the wave function are neatly captured respectively by the position-dependent (conditional) average and variance of the epistemically restricted momentum fluctuation. They are in turn respectively determined by the real and imaginary parts of the weak momentum value, permitting a reconstruction of wave function using weak momentum value measurement, and an interpretation of momentum weak value in term of epistemically restricted momentum fluctuations. The ERPS representation thus provides a transparent and rich framework to study the deep conceptual links between quantum uncertainty embodied in epistemic restriction, quantum wave function, and weak momentum measurement with position post-selection, which may offer useful insight to better understand their meaning.  
  
\end{abstract}    

\maketitle                   

\section{introduction}

It is remarkable that ever since the inception of quantum mechanics, and despite its uncontested pragmatical successes, physicists still have no consensus on the meaning of pure quantum state or wave function, the key element of the theory. Is wave function a real-physical thing independent of measurement, or is it a mathematical tool which represents some form of information? A better intuition about wave function may offer fresh insight into a deeper understanding of quantum paradoxes \cite{Aharonov-Daniel book}, and may hold the key to identify the elusive physical resources underlying the power of quantum information protocols relative to their classical counterparts \cite{Howard contextuality magic,Veitch negativity as resource,Vedral quantum computational resource}. A powerful and rich method which could aid our (classical) intuition to grasp the quantum states and is useful to asses the quantum-classical correpondence and contrast, is by mapping the quantum states onto quasiprobability distributions over phase space \cite{Wigner paper on Wigner function,Hillery quasiprobability review,Lee quasiprobability review,Ferrie review quasiprobability}. Quasiprobability distributions are the quantum mechanical `analogue' of classical phase space distribution. Wigner function, the earliest and the most well-known quasiprobability distribution, satisfies most of the intuitive requirements for a `proper' probability distribution over classical phase space, but it may take on negative values. Despite of this, Wigner function can be operationally determined via standard strong (projective) measurements, leading to a method to reconstruct the underlying quantum state \cite{Bertrand quantum state tomography,Vogel quantum state tomography,Smithey quantum state tomography,Gibbons discrete Wigner function,Leonhardt discrete Wigner function}. The intuition and insight provided by the quasiprobability representation has recently led to an important application in the field of quantum computation to devise an efficient classical simulation and estimation of certain class of quantum computational algorithms \cite{Veitch efficient simulation nonnegative WF 2,Mari-Eisert classical simulation,Rahemi-Keshari efficient simulation quantum optics,Pashayan efficient estimation nonnegative quasiprob}. 

However, it is still not fully understood, how, two most distinctive features of microscopic world, namely quantum uncertainty and entanglement, are deeply and transparently manifested in the quasiprobability phase space representation. In addition, the mathematical structure of complex quantum wave function is not transparently reflected in the associated quasiprobability distributions. The construction of quasiprobability is formal, guided more by ingenious mathematical tricks, rather than coming from a deep thinking about quantum uncertainty and entanglement \cite{Hillery quasiprobability review,Lee quasiprobability review,Ferrie review quasiprobability}. Moreover, there are infinitely many quasiprobability representations, and the choice seems arbitrary. Can an insightful phase space representation of quantum mechanics be singled out or motivated uniquely by requiring the microscopic world to obey the quantum uncertainty relation and to transparently reflect the structure of quantum wave function. Partly motivated by this conceptual problem, a novel phase space representation for quantum mechanics was proposed in Ref. \cite{Agung-Daniel model}. In the phase space representation, a quantum wave function is associated with a phase space distribution which explicitly incorporates a specific epistemic restriction parameterized by a global random variable on the order of Planck constant, transparently manifesting quantum uncertainty relation in a classical phase space.   

Meanwhile, in the last decades, there have been a lot of interest in a novel and intriguing concept of weak value measurement over pre- and post-selected ensembles \cite{Aharonov-Daniel book,Aharonov weak value,Jozsa complex weak value,Lundeen complex weak value}. To an extent, this measurement protocol reflects our ordinary intuition about measurement in everyday life \cite{Wiseman Bohmian mechanics}, wherein one first gently measures a physical quantity without much disturbing the system and then followed by a post-selection on a sub-ensemble of interest. Yet, the outcomes of weak value measurement may go beyond the range of values obtained via the standard strong measurement, fuelling hot debates about its meaning \cite{Svensson meaning weak value,Ferrie statistical weak value,Sokolovski negative probability,Anomalous weak value - contextuality,Johansen weak value as Bayesian efficient estimator,Hall prior information,Karanjai weak value in epistricted model,Vaidman review meaning weak value}. Weak value measurement offers an access into a rich microscopic regime unreachable by the standard value, and has found many interesting applications in quantum metrology \cite{Hosten weak value quantum metrology,Dixon weak value quantum metrology}, to study the conundra in quantum foundation \cite{Aharonov-Daniel book,Lundeen weak value Hardy's paradox,Wiseman weak valued distribution momentum transfer,Steinberg average trajectory,Aharonov three boxes paradox,Resch three boxes paradox,Aharonov Cheshire cat,Denkmayr Cheshire cat}, and recently for measuring the quantum state directly \cite{Lundeen direct measurement of the quantum state,Salvail direct measurement of quantum state,Malik direct measurement of quantum state,Mirhosseini direct measurement of quantum state}.  

In the present work, we show that the above two seemingly different concepts, i.e., the `epistemically restricted phase space' (ERPS) representation \cite{Agung-Daniel model} and weak value measurement \cite{Aharonov weak value} are deeply interrelated, whose clarification may offer fresh insight into their meaning. Namely, first, the ERPS distribution can be defined operationally using weak measurement of momentum followed by post-selection on position, called as weak momentum value. Moreover, in the ERPS representation, the phase and amplitude of the wave function are directly captured respectively by the position-dependent (conditional) average and variance of the epistemically (statistically) restricted momentum fluctuation, which are, in turn, respectively determined by the real and imaginary parts of the weak momentum value. This observation permits a simple reconstruction of the wave function via the weak momentum value measurement, and offers an interpretation of the complex weak momentum value in term of epistemically restricted momentum fluctuation. We further speculate on the possible deep relations between the epistemically restricted momentum field, Wiseman's naively observable average momentum field \cite{Wiseman Bohmian mechanics}, and Hall-Johansen's best estimation of momentum given position \cite{Johansen weak value as Bayesian efficient estimator,Hall best estimate,Hall prior information}, mediated operationally by the weak momentum measurement with position post-selection. We expect that these deep links between different fundamental concepts may shed new light on the elusive meaning of wave function, to be further elaborated in the future \cite{Agung epistemic interpretation}.  

\section{Epistemically restricted phase space representation for quantum mechanics} 

Consider a general system of $N$ spatial degrees of freedom with a configuration $q=(q_1,\dots,q_N)$. First, given a preparation characterized by a pure quantum state $\ket{\psi}$, we write the associated wave function $\psi(q)\doteq\braket{q|\psi}$ in polar form as 
\begin{eqnarray}
\psi(q)=\sqrt{\rho(q)}e^{iS(q)/\hbar},
\label{wave function}
\end{eqnarray} 
so that the amplitude $\rho(q)$ and the phase $S(q)$, both are real-valued functions, are given by $\rho(q)=|\psi(q)|^2$ and $S(q)=\frac{\hbar}{2i}\big(\ln\psi(q)-\ln\psi^*(q)\big)$. Following Born, we interpret $\rho(q)$ as the probability density that the system has a configuration $q$. We then define a conditional probability of the momentum $p=(p_1,\dots,p_N)$ given the conjugate positions $q$, associated with the wave function $\psi(q)$ via its phase and amplitude, as  \cite{Agung-Daniel model}
\begin{eqnarray}
{\mathbb P}_{\psi}(p|q,\xi)=\prod_{n=1}^N\delta\Big(p_n-\Big(\partial_{q_n}S +\frac{\xi}{2}\frac{\partial_{q_n}\rho}{\rho}\Big)\Big).
\label{fundamental epistemic restriction}
\end{eqnarray}
Here $\xi$ is a global-nonseparable variable with dimension of action, fluctuating randomly on a microscopic time scale with a probability density $\chi(\xi)$, so that its average and variance are constant in space and time, given by 
\begin{eqnarray}
\overline{\xi}\doteq\int {\rm d}\xi~\xi~\chi(\xi)=0~~~\&~~~\overline{\xi^2}=\hbar^2.
\label{Planck constant}
\end{eqnarray}

The conditional probability distribution of phase space associated with a preparation characterized by a quantum state $\ket{\psi}$ thus reads 
\begin{eqnarray}
{\mathbb P}_{\psi}(p,q|\xi)&=&{\mathbb P}_{\psi}(p|q,\xi)\rho(q)\nonumber\\
&=&\prod_{n=1}^N\delta\Big(p_n-\Big(\partial_{q_n}S +\frac{\xi}{2}\frac{\partial_{q_n}\rho}{\rho}\Big)\Big)\rho(q).
\label{restricted phase space distribution}
\end{eqnarray}
Clearly, unlike Wigner function, the phase space distribution ${\mathbb P}_{\psi}(p,q|\xi)$ is, by construction, always nonnegative. One can then show that the quantum expectation value of any quantum observable $\hat{O}$ up to second order in momentum operator $\hat{p}$ over a quantum state $\ket{\psi}$ can be expressed as the conventional statistical average of a classical quantity $O(p,q)$ over the phase space distribution of Eq. (\ref{restricted phase space distribution}), i.e.,  \cite{Agung-Daniel model}:
\begin{eqnarray}
\braket{\psi|\hat{O}|\psi}=\int{\rm d}q{\rm d}\xi{\rm d}p~O(p,q){\mathbb P}_{\psi}(p,q|\xi)\chi(\xi)\doteq\braket{O}_{\psi}. 
\label{equivalence principle}
\end{eqnarray} 
Here, the Hermitian operator $\hat{O}$ must take the same form as that obtained by applying the standard canonical quantization scheme to $O(p,q)$ with a specific ordering of operators. Moreover, for $O(p,q)$ with cross terms between momentum of different degrees of freedom, i.e. $p_ip_j$, $i\neq j$, the nonseparability of $\xi$ is indeed indispensable.  See Section Methods in Ref. \cite{Agung-Daniel model} for a proof.     

Reading Eq. (\ref{equivalence principle}) from the right hand side to the left, i.e., as a reconstruction of quantum expectation value from classical statistical phase space average, one can transparently see that the form of Hermitian quantum observable $\hat{O}$, including its operator ordering, is mathematically related to the definition of wave function $\psi$ given in Eq. (\ref{wave function}). As an example of operator ordering, consider a system with one spatial dimension and a general classical physical quantity up to second order in momentum: $O(p,q)=A(q)p^2+B(q)p+C(q)$, where $A(q)$, $B(q)$ and $C(q)$ are real-valued functions of $q$. Then, it is straightforward to show that, imposing the equality of Eq. (\ref{equivalence principle}), the correspondence rule between the classical quantity $O(p,q)$ and the associated Hermitian operator $\hat{O}$ takes the following ordering: $O(p,q)=A(q)p^2+B(q)p+C(q)\mapsto\hat{p}A(\hat{q})\hat{p}+\frac{1}{2}(\hat{p}B(\hat{q})+B(\hat{q})\hat{p})+C(\hat{q})\doteq\hat{O}$ (see Section Method in \cite{Agung-Daniel model}). It is however not clear how the ordering of operators in the above phase space representation is related to the different orderings of operators in various quasiprobability representations \cite{Hillery quasiprobability review,Lee quasiprobability review}. 

We note that while the above proposed phase space distribution and the so-called Q (or more general, Husimi) quasiprobability distribution function are both non-negative for arbitrary wave functions, the correspondence rule between the quantum observable and the associated classical quantity in the two phase space representations are different. To see this, consider a simple example where the quantum observable has the form $\hat{O}=\hat{q}^2$. Then, within our phase space representation, the associated classical quantity takes the form $O=q^2$, as in Wigner-Weyl correspondence. On the other hand, within the Q phase space formalism, the associated classical quantity takes the form $O=q^2-\hbar/2$ \cite{Lee quasiprobability review}. In this sense, the ERPS representation thus inherits the desirable properties of Wigner and Q phase space representations: namely, the ERPS distribution is non-negative as Q function, and the correspondence rule between the Hermitian operator and classical quantity is intuitive as in Wigner-Weyl correspondence rule.

A few further notes on the physical interpretation of the phase space representation are in order. First, the delta functional form of the conditional probability of Eq. (\ref{fundamental epistemic restriction}) enables us to write $p$ explicitly as a function of $(q,\xi)$ as  
\begin{eqnarray}
p_n(q;\xi,\psi)=\partial_{q_n}S+\frac{\xi}{2}\frac{\partial_{q_n}\rho}{\rho},
\label{epistricted random momentum field}
\end{eqnarray}
$n=1,\dots, N$. It describes a momentum field fluctuating randomly due to the fluctuation of $\xi$. As argued in Ref. \cite{Agung-Daniel model}, Eq. (\ref{epistricted random momentum field}) can be interpreted as a fundamental epistemic/statistical restriction \cite{Spekkens toy model with epistemic restriction,Bartlett reconstruction of Gaussian QM with epistemic restriction,Hardy epistricted model,Emerson PhD thesis} on the allowed ensemble of trajectories. Namely, while in conventional classical statistical mechanics we can prepare an ensemble of trajectories with a desired distribution of positions $\rho(q)$ using arbitrary momentum field, in the above phase space model, the allowed form of momentum field must depend on the targeted $\rho(q)$ as prescribed by Eq. (\ref{epistricted random momentum field}). Conversely, given a momentum field $p(q;\xi,\psi)$, unlike in conventional classical statistical mechanics, it is no longer granted to assign each trajectory in the momentum field an arbitrary weight $\rho(q)$. Hence, the allowed forms of distribution of positions  $\rho(q)$ depends on, thus is restricted fundamentally by, the underlying momentum field, satisfying Eq. (\ref{epistricted random momentum field}).   For this reason, we shall refer to ${\mathbb P}_{\psi}(p,q|\xi)$ of Eq. (\ref{restricted phase space distribution}) as the `epistemically restricted phase space (ERPS) distribution' associated with the wave function $\psi(q)$, and $p(q;\xi,\psi)$ defined in Eq. (\ref{epistricted random momentum field}) as the epistemically restricted momentum field. 

We have also argued in Ref \cite{Agung-Daniel model} that the epistemic restriction of Eqs. (\ref{fundamental epistemic restriction}) or (\ref{epistricted random momentum field}), together with Eq. (\ref{Planck constant}), transparently embodies the quantum uncertainty relation in classical phase space. For example, Eqs. (\ref{epistricted random momentum field}) and (\ref{Planck constant}) directly implies that the standard deviations of position and momentum, respectively denoted by $\sigma_q$ and $\sigma_p$, must satisfy the Heisenberg-Kennard uncertainty relation $\sigma_q\sigma_p\ge\hbar/2$. In this sense, the epistemic restriction of Eq. (\ref{epistricted random momentum field}) could be regarded as a `local', i.e. position-dependent, thus `stronger' manifestation of quantum uncertainty relation in classical phase space. Moreover, the unitary quantum dynamics, namely the Schr\"odinger equation, is uniquely singled out by requiring the statistically restricted ensemble of trajectories satisfying Eqs. (\ref{epistricted random momentum field}) and (\ref{Planck constant}) to further respect the conservation of average energy and trajectories (probability current) \cite{Agung-Daniel model}. This includes those that describe quantum dynamical interactions between subsystems, generating quantum entanglement. 

As an example of the ERPS representation, consider a spatially one dimensional system prepared in a Gaussian wave function: $\psi_{\rm G}(q)=(\frac{1}{2\pi\sigma_q^2})^{\frac{1}{4}}\exp(-\frac{(q-q_o)^2}{4\sigma_q^2}+\frac{i}{\hbar}p_oq)$. How does the ERPS distribution look like? First, as per Eq. (\ref{wave function}), we have $S_{\rm G}(q)=p_oq$ and $\rho_{\rm G}(q)=(\frac{1}{2\pi\sigma_q^2})^{\frac{1}{2}}\exp(-\frac{(q-q_o)^2}{2\sigma_q^2})$. Inserting into Eq. (\ref{restricted phase space distribution}), the ERPS distribution takes the form ${\mathbb P}_{\psi_{\rm G}}(p,q|\xi)=\delta\big(p-p_o+\frac{\xi}{2}\frac{(q-q_o)}{\sigma_q^2}\big)(\frac{1}{2\pi\sigma_q^2})^{\frac{1}{2}}\exp(-\frac{(q-q_o)^2}{2\sigma_q^2})$. Note that the marginal distribution of position is just given by a Gaussian distribution, i.e.: ${\mathbb P}_{\psi_{\rm G}}(q)=\int{\rm d}p{\rm d}\xi{\mathbb P}_{\psi_{\rm G}}(p,q|\xi)\chi(\xi)=\rho_{\rm G}(q)=(\frac{1}{2\pi\sigma_q^2})^{\frac{1}{2}}\exp(-\frac{(q-q_o)^2}{2\sigma_q^2})$. On the other hand, computing the marginal distribution of momentum, assuming that the distribution of $\xi$ has the form $\chi(\xi)=\frac{1}{2}\delta(\xi-\hbar)+\frac{1}{2}\delta(\xi+\hbar)$ satisfying Eq. (\ref{Planck constant}), we straightforwardly obtain a Gaussian distribution of momentum, i.e.: ${\mathbb P}_{\psi_{\rm G}}(p)=\int{\rm d}\xi{\rm d}q{\mathbb P}_{\psi_{\rm G}}(p,q|\xi)\chi(\xi)=\int{\rm d}\xi(\frac{2\sigma_q^2}{\pi\xi^2})^{\frac{1}{2}}\exp(-\frac{2\sigma_q^2}{\xi^2}(p-p_o)^2)\chi(\xi)=(\frac{2\sigma_q^2}{\pi\hbar^2})^{\frac{1}{2}}\exp(-\frac{2\sigma_q^2}{\hbar^2}(p-p_o)^2)$, with the standard deviation $\sigma_p^2=\hbar^2/(4\sigma_q^2)$. Hence, for Gaussian wave functions, both the marginal distributions of position and momentum are equal to the corresponding quantum probabilities obtained in standard (strong) measurement, i.e.: ${\mathbb P}_{\psi_{\rm G}}(q)=|\braket{q|\psi_{\rm G}}|^2$ and ${\mathbb P}_{\psi_{\rm G}}(p)=|\braket{p|\psi_{\rm G}}|^2$, as for the Wigner function. Note however that, unlike the Wigner function, this result cannot be extended to general wave functions. Namely, while we always have ${\mathbb P}_{\psi}(q)=\rho(q)=|\braket{q|\psi}|^2$ for general wave function, in general we have ${\mathbb P}_{\psi}(p)\neq|\braket{p|\psi}|^2$, as the case for e.g. Q-function \cite{Lee quasiprobability review}. Hence, $p$ in ${\mathbb P}_{\psi}(p,q|\xi)$ is not in general equal to the outcome of momentum measurement.    

Further, unlike general quasiprobability representations which, by construction, are devised to express the quantum expectation value of arbitrary Hermitian operators into phase space integration evocative of classical statistical average at the cost of allowing negative quasiprobability \cite{Spekkens negativity-contextuality,Ferrie theorem negative quasiprobability}, in the ERPS representation, Eq. (\ref{equivalence principle}) applies only for Hermitian operators $\hat{O}$ up to second order in momentum operator. Moreover, note that quasiprobability distributions are defined (bi)linearly in term of wave function. By contrast, as explicitly seen in Eq. (\ref{restricted phase space distribution}), the ERPS distribution ${\mathbb P}_{\psi}(p,q|\xi)$ is obtained by nonlinearly mapping the wave function $\psi$ \cite{Montina linearity}. In this sense, we have thus given in some requirements of the quasiprobability representations and trading them with the following main conceptual advantages of ERPS representation: the quantum uncertainty relation is transparently manifested in the form of epistemic restriction in classical phase space, the ERPS distributions associated with wave functions are non-negative and the classical quantities associated with quantum observables have intuitive forms, and the Schr\"odinger equation can be directly derived by imposing the conservation of average energy and probability current. We expect that, like quasiprobability representation, ERPS representation may offer new insight to devise efficient classical simulation and/or estimation of certain class of quantum computational algorithms \cite{Veitch efficient simulation nonnegative WF 2,Mari-Eisert classical simulation,Rahemi-Keshari efficient simulation quantum optics,Pashayan efficient estimation nonnegative quasiprob}. An effort along this direction is reported in a different work \cite{ER simulating quantum computation}.

Besides satisfying all the axioms for a true probability, the ERPS distribution of Eq. (\ref{restricted phase space distribution}) also satisfies the following two intuitive requirements. First, consider a composite of two subsystems with a configuration $q=(q_1,q_2)$ and the conjugate momentum $p=(p_1,p_2)$. Assume that the preparations of the two subsystems are independent of each other so that the total wave function of the composite is factorizable, $\psi(q_1,q_2)=\psi_1(q_1)\psi_2(q_2)$. Noting Eq. (\ref{wave function}), in this case, the total phase of the composite wave function is decomposable $S(q_1,q_2)=S_1(q_1)+S_2(q_2)$, and the amplitude is factorizable $\rho(q_1,q_2)=\rho(q_1)\rho(q_2)$. Inserting these into Eq. (\ref{restricted phase space distribution}), the ERPS distribution associated with a pair of independent preparations is thus conditionally separable, i.e., ${\rm P}_{\psi}(p,q|\xi)={\rm P}_{\psi}(p|q,\xi)\rho(q)={\rm P}_{\psi_1}(p_1,q_1|\xi){\rm P}_{\psi_2}(p_2,q_2|\xi)$. Note however that, due to the nonseparability of $\xi$, the distribution of $(p,q,\xi)$ is nonfactorizable: ${\rm P}_{\psi}(p,q,\xi)={\rm P}_{\psi}(p,q|\xi)\chi(\xi)\neq{\rm P}_{\psi_1}(p_1,q_1,\xi){\rm P}_{\psi_2}(p_2,q_2,\xi)$. Next, let us apply the usual unitary phase space shift operator, $\ket{\psi}\mapsto\hat{U}_D\ket{\psi}$, where $\hat{U}_D=e^{\frac{i}{\hbar}q_0\hat{p}-\frac{i}{\hbar}p_0\hat{q}}$. In this case, the wave function transforms as $\psi(q)\mapsto\psi(q-q_0)e^{-ip_0(q-q_0)/\hbar-ip_0q_0/2\hbar}$. Noting Eq. (\ref{wave function}), this means that the phase transforms as $S(q)\mapsto S(q-q_0)-p_0q+p_0q_0/2$ and the amplitude as $\rho(q)\mapsto\rho(q-q_0)$. Inserting into Eq. (\ref{restricted phase space distribution}), the ERPS distribution thus transforms as ${\rm P}_{\psi}(p,q|\xi)\mapsto{\rm P}_{\psi}(p-p_0,q-q_0|\xi)$, hence, it is covariant under the phase space shift transformation. 

Finally, since we work within the conventional statistical theory, the ERPS representation for pure states (wave functions) discussed above can be naturally extended to incoherent mixture (convex combination) of pure states. For illustration, consider a mixed state $\hat{\varrho}=\sum_{l=1}^Lr_l|\psi_l\rangle\langle\psi_l|$, where $0\le r_l\le 1$, $\sum_{l=1}^Lr_l=1$. Then, the ERPS distribution associated with $\hat{\varrho}$ must take the form ${\mathbb P}_{\hat{\varrho}}(p,q|\xi)\doteq\sum_{l=1}^Lr_l{\mathbb P}_{\psi_l}(p,q|\xi)$, where each ${\mathbb P}_{\psi_l}(p,q|\xi)$ is given by Eq. (\ref{restricted phase space distribution}). The conventional statistical formula to compute the average value of Eq. (\ref{equivalence principle}) still applies: we only need to use ${\mathbb P}_{\hat{\varrho}}(p,q|\xi)$ in place of ${\mathbb P}_{\psi}(p,q|\xi)$. Namely, we have ${\rm Tr}\{\hat{\varrho}\hat{O}\}=\sum_lr_l\braket{\psi_l|\hat{O}|\psi_l}=\sum_lr_l\int{\rm d}q{\rm d}\xi{\rm d}p~O(p,q){\mathbb P}_{\psi_l}(p,q|\xi)\chi(\xi)=\int{\rm d}q{\rm d}\xi{\rm d}p~O(p,q)\sum_lr_l{\mathbb P}_{\psi_l}(p,q|\xi)\chi(\xi)=\int{\rm d}q{\rm d}\xi{\rm d}p~O(p,q){\mathbb P}_{\hat{\varrho}}(p,q|\xi)\chi(\xi)\doteq\braket{O}_{\hat{\varrho}}$.  
 
\section{ERPS distribution from weak measurement of momentum with post-selection on conjugate position}

We have nonlinearly mapped the wave function to obtain the ERPS distribution of Eq. (\ref{restricted phase space distribution}). This raises a question whether such a nonlinear mapping can be implemented experimentally. We show in this section that the ERPS distribution can indeed be defined operationally using weak value measurement \cite{Aharonov weak value}. For simplicity, consider a system of one spatial dimension; assume that it is prepared in a pure quantum state $\ket{\psi}$. Suppose we make a weak momentum measurement and then followed by post-selection on a sub-ensemble passing through a position $q$ implemented by making a strong projective measurement onto $\ket{q}\bra{q}$. We thereby obtain the complex weak momentum value at $q$, denoted by $p^{\rm w}(q;\psi)$, as: 
\begin{eqnarray}
p^{\rm w}(q;\psi)\doteq \frac{\braket{q|\hat{p}|\psi}}{\braket{q|\psi}}. 
\label{weak value momentum}
\end{eqnarray}
Below we shall refer to $p^{\rm w}(q;\psi)$ simply as weak momentum value. Writing the wave function in polar form as in Eq. (\ref{wave function}), the real and imaginary parts of the weak momentum value are straightforwardly given by 
\begin{eqnarray}
{\rm Re}\big\{p^{\rm w}(q;\psi)\big\}=\partial_qS~~~\&~~~{\rm Im}\big\{p^{\rm w}(q;\psi)\big\}=-\frac{\hbar}{2}\frac{\partial_q\rho}{\rho}. 
\label{real and imaginary weak value momentum}
\end{eqnarray}
It has been shown in general that both the real and imaginary parts of the weak value can be inferred respectively from the `average' shift of the position and momentum of the measuring device pointer \cite{Jozsa complex weak value,Lundeen complex weak value}. 

Noting Eq. (\ref{real and imaginary weak value momentum}), the epistemically restricted random momentum field of Eq. (\ref{epistricted random momentum field}) can thus be written in term of the weak momentum value as 
\begin{eqnarray}
p(q;\xi,\psi)={\rm Re}\big\{p^{\rm w}(q;\psi)\big\}-\frac{\xi}{\hbar}{\rm Im}\big\{p^{\rm w}(q;\psi)\big\}.
\label{random momentum field as weak momentum values}
\end{eqnarray}
Equation (\ref{random momentum field as weak momentum values}) can immediately be extended to $N$ degrees of freedom so that the conditional distribution of momentum of Eq. (\ref{fundamental epistemic restriction}) can be written as ${\mathbb P}_{\psi}(p|q,\xi)=\prod_{n=1}^{N}\delta\big(p_n-\big[{\rm Re}\big\{p^{\rm w}_n(q;\psi)\big\}-\frac{\xi}{\hbar}{\rm Im}\big\{p^{\rm w}_n(q;\psi)\big\}\big]\big)$. The ERPS distribution thus reads, in term of the weak momentum value, as 
\begin{eqnarray}
&&{\mathbb P}_{\psi}(p,q|\xi)={\mathbb P}_{\psi}(p|q,\xi)\rho(q)\nonumber\\
&=&\prod_{n=1}^{N}\delta\Big(p_n-\big[{\rm Re}\big\{p^{\rm w}_n(q;\psi)\big\}-\frac{\xi}{\hbar}{\rm Im}\big\{p^{\rm w}_n(q;\psi)\big\}\big]\Big)\rho(q).
\end{eqnarray}
Hence, the ERPS distribution can indeed be operationally obtained using weak measurement of momentum followed by post-selection on the conjugate position, combined with the introduction of the (hypothetical) global random variable $\xi$ satisfying Eq. (\ref{Planck constant}). It is thus more than just a mathematical artefact. Such a weak momentum value measurement has already been performed as reported in Ref. \cite{Steinberg average trajectory}. 

The above observation conversely suggests the following statistical interpretation of the meaning of complex weak momentum value within the ERPS representation. For simplicity, below we work again in one spatial dimension. First, in the ERPS representation, from Eqs. (\ref{random momentum field as weak momentum values}) and (\ref{Planck constant}), the real part of the weak momentum value is equal to the position-dependent (conditional) average of the epistemically restricted random momentum field $p(q;\xi,\psi)$ of the system over the fluctuation of $\xi$, i.e.:
\begin{eqnarray}
\overline{p}(q;\psi)&\doteq&\int{\rm d}\xi~p(q;\xi,\psi)\chi(\xi)=\int{\rm d}\xi{\rm d}p~p~{\mathbb P}_{\psi}(p|q,\xi)\chi(\xi)\nonumber\\
&=&\partial_qS(q)={\rm Re}\big\{p^{\rm w}(q;\psi)\big\}, 
\label{the real part of the weak value momentum}
\end{eqnarray}
where, in the last equality we have used Eq. (\ref{real and imaginary weak value momentum}).  On the other hand, the square of the imaginary part of the weak momentum value is equal to the position-dependent variance of the restricted random momentum field of the system, i.e.: 
\begin{eqnarray}
\overline{\overline{p}}(q;\psi)&\doteq&\int{\rm d}\xi\big(p(q;\xi,\psi)-\overline{p}(q;\psi)\big)^2\chi(\xi)\nonumber\\
&=&\int{\rm d}\xi{\rm d}p\big(p-\partial_qS\big)^2{\mathbb P}_{\psi}(p|q,\xi)\chi(\xi)\nonumber\\
&=&\frac{\hbar^2}{4}\Big(\frac{\partial_q\rho}{\rho}\Big)^2=\big({\rm Im}\big\{p^{\rm w}(q;\psi)\big\}\big)^2. 
\label{the imaginary part of the weak value momentum}
\end{eqnarray} 
Here, to get the second equality we have used (\ref{the real part of the weak value momentum}), the third is due to Eqs. (\ref{epistricted random momentum field}) and (\ref{Planck constant}), and the last equality is implied by Eq. (\ref{real and imaginary weak value momentum}). We note that a different statistical interpretation of weak value, within a classical model with an epistemic restriction taking the form of Heisenberg uncertainty relation reproducing Gaussian quantum mechanics \cite{Bartlett reconstruction of Gaussian QM with epistemic restriction}, is reported in Ref. \cite{Karanjai weak value in epistricted model}. 

To summarize, the operational protocol of weak momentum measurement with post-selection on the conjugate position, combined with the introduction of a global random variable $\xi$ satisfying Eq. (\ref{Planck constant}), defines the ERPS distribution ${\mathbb P}_{\psi}(p,q|\xi)$, in a specific way. Moreover, the real part of the weak momentum value is manifested in the position-dependent (conditional) average of the statistically restricted momentum fluctuation ${\overline{p}}(q;\psi)$ as in Eq. (\ref{the real part of the weak value momentum}), and up to its sign, the imaginary part of weak momentum value is manifested in the variance of the the statistically restricted momentum fluctuation $\overline{\overline{p}}(q;\psi)$ as in (\ref{the imaginary part of the weak value momentum}), offering a statistical meaning of complex weak momentum value. In particular, the above phase space picture naturally leads to a conjecture that, the randomness in each single repetition of weak momentum measurement is due to the random fluctuation of the epistemically restricted momentum field of Eq. (\ref{epistricted random momentum field}), with an average and a variance that correspond respectively to the real and imaginary parts of the weak momentum value (see also next section on Wiseman's naive scheme to observe average momentum and its relation with weak momentum value). This conjecture can be checked in experiment if we could probe the outcome of each single shot of weak measurement.  Another interesting important question is that whether a phase space distribution with an epistemic restriction can be singled out operationally via weak momentum value, uniquely, by imposing some further physically intuitive requirements, such as separability for independent preparations and covariance under phase space shift transformation discussed at the end of the previous section. 

\section{Discussion: wave function tomography, Wiseman's average momentum field, and best estimation of momentum given information on position}

Equation (\ref{real and imaginary weak value momentum}) suggests a simple method for the reconstruction of quantum wave function  $\psi(q)=\sqrt{\rho(q)}\exp(iS(q)/\hbar)$ from weak measurement of momentum with post-selection on the conjugate position. First, integrating the left equation in (\ref{real and imaginary weak value momentum}), we obtain the phase of the associated wave function up to a constant or a global phase as: $S(q)=\int^q{\rm d}q'{\rm Re}\big\{p^{\rm w}(q';\psi)\big\}$. On the other hand, the amplitude of the original wave function can be computed by integrating the right equation in (\ref{real and imaginary weak value momentum}) as: $\rho(q)=\mathcal{C}\exp\Big(-\frac{2}{\hbar}\int^q{\rm d}q'{\rm Im}\big\{p^{\rm w}(q';\psi)\big\}\Big)$, where $\mathcal{C}$ is a normalization constant. Moreover, noting Eqs. (\ref{the real part of the weak value momentum}) and (\ref{the imaginary part of the weak value momentum}), the reconstructed phase and amplitude of the wave function, thus the whole quantum information concealed in a wave function, are respectively captured neatly by the position-dependent (conditional) average and variance of the statistically restricted momentum field defined in Eq. (\ref{epistricted random momentum field}). The ERPS representation thus reveals a deep conceptual link between the epistemic restriction transparently manifesting quantum uncertainty relation in phase space, and the abstract mathematical structure of complex quantum wave function, mediated by the operationally well-defined notion of weak measurement of momentum followed by post-selection on the conjugate position. This conceptual link is further elaborated in a different work to devise an epistemic interpretation of quantum wave function and quantum uncertainty \cite{Agung epistemic interpretation}; see the last paragraph of this section. 

Note that in the scheme for a direct measurement of wave function  reported in \cite{Lundeen direct measurement of the quantum state}, i.e., by weakly measuring $\ket{q}\bra{q}$ followed by a post-selection on a subensemble with vanishing momentum (via strong momentum measurement), the real and imaginary parts of the associated weak value correspond directly to the real and imaginary parts of the quantum wave function, whose physical interpretation are not easy to grasp. By contrast, in our reconstruction scheme above, the real and imaginary parts of the weak momentum value correspond directly to the phase and the amplitude of the wave function, whose physical interpretation are more transparent. In Ref. \cite{Bamber direct weak measurement of Dirac distribution}, it is shown that Dirac-Kirkwood quasiprobability distribution \cite{Dirac quasiprobability distribution,Kirkwood quasiprobability distribution,Lee quasiprobability review} can be defined operationally using weak measurement of $\ket{q}\bra{q}$ followed by a strong projection over $\ket{p}\bra{p}$ (without post-selection). However, while Dirac-Kirkwood quasiprobability distribution appears formally to satisfy the Bayes rule \cite{Steinberg Bayes rules for complex conditional probability,Hofmann Bayes rules for complex conditional probability}, its transparent interpretation is hampered by the fact that it is complex-valued. Moreover, unlike the ERPS distribution, its relation with quantum uncertainty relation is not immediate.    

Next, in Ref. \cite{Wiseman Bohmian mechanics}, Wiseman considered an intuitive but seemingly `naive' definition of position-dependent (conditional) average momentum, i.e., in a way that would make sense to classical physicists, and interpreted it operationally in term of weak value measurement. Namely, one makes weak momentum measurement followed by position post selection, yielding random value, and takes the average over an infinite repetitions of such measurement procedure. He then showed that the average momentum field thus defined, is equal to the real part of the weak momentum value given by the left equation in (\ref{real and imaginary weak value momentum}). For example, for a particle of mass $m$ in a scalar potential, we have $ m v_{\rm W}\doteq p_{\rm W}(q)={\rm Re}\{p^{\rm w}(q;\psi)\}=\partial_qS(q)$, where $p_{\rm W}(q)$ and $v_{\rm W}(q)$ are the Wiseman's average momentum and velocity fields. This can be integrated in time to construct `average trajectories' which in turn coincides with the Bohmian trajectories. Such `average trajectories', which was calculated numerically for the first time in Ref. \cite{numerical Bohmian trajectories} in the context of Bohmian mechanics, has been observed by Steinberg's group in a beautiful double-slit experiment reported in Ref. \cite{Steinberg average trajectory}. Within the ERPS representation, as shown in (\ref{the real part of the weak value momentum}), Wiseman's naively observable average momentum field is just the position-dependent (conditional) average of the statistically restricted momentum field defined in Eq. (\ref{epistricted random momentum field}) over the fluctuations of $\xi$; i.e., we have $p_{\rm W}(q)={\rm Re}\big\{\frac{\braket{q|\hat{p}|\psi}}{\braket{q|\psi}}\big\}=\partial_qS(q)=\overline{p}(q;\psi)$. Our interpretation of weak momentum value is thus very similar to that of Wiseman's. Moreover, as also stated at the end of the previous section, we may naturally speculate that the randomness in each single repetition of weak momentum measurement in Wiseman's `naive scheme' is due to the random fluctuation of $\xi$ in Eq. (\ref{epistricted random momentum field}). Note however that Wiseman's starting point is standard quantum mechanics, while we started from a statistical phase space model to reconstruct quantum mechanics from scratch using the notion of epistemic restriction parameterized by a global variable $\xi$ fluctuating randomly on the order of Planck constant \cite{Agung-Daniel model}.  

Further, in Ref. \cite{Hall best estimate}, working within the standard formalism of quantum mechanics, Hall argued that $\partial_qS(q)={\rm Re}\{\frac{\braket{q|\hat{p}|\psi}}{\braket{q|\psi}}\}$ can be interpreted as the `best classical estimate' of quantum momentum compatible with the conjugate position $q$, minimizing a suitably defined quantum mean-squared error. This idea has been further discussed by Johansen in Ref. \cite{Johansen weak value as Bayesian efficient estimator} to interpret weak value in term of theory of best estimation, and is largely expanded by Hall in Ref. \cite{Hall prior information}. Noting this, within the ERPS representation, we might interpret the decomposition of momentum field in Eq. (\ref{epistricted random momentum field}) epistemically (as opposed to physical decomposition) as follows. Assume that a preparation characterized by a wave function $\psi(q)=\sqrt{\rho(q)}e^{\frac{i}{\hbar}S(q)}$ generates a random momentum field $p(q;\xi,\psi)$. Then, given information on the conjugate position $q$, the first term on the right hand side of the decomposition in Eq. (\ref{epistricted random momentum field}), i.e., $\partial_qS(q)$, is interpreted as the best estimate of the momentum field, and the second term $\frac{\xi}{2}\frac{\partial_q\rho(q)}{\rho(q)}=p(q;\xi,\psi)-\partial_qS(q)$ as the estimation error. It is interesting to ask if this interpretation could link the Cramer-Rao inequality limiting an unbiased estimation \cite{Papoulis and Pillai book} to the Heisenberg quantum uncertainty relation. If this epistemic interpretation of the decomposition of Eq. (\ref{epistricted random momentum field}) is tenable, along with the Copenhagen spirit, we might argue within the ERPS-model of Ref. \cite{Agung-Daniel model} that quantum wave function $\psi(q)=\sqrt{\rho(q)}e^{\frac{i}{\hbar}S(q)}$ does not represent an agent(observer)-independent objective physical reality \cite{PBR theorem}, but a mathematical tool which conveniently summarizes the agent's best estimate of momentum given information on the conjugate position, in the presence of an epistemic restriction. This idea is elaborated in detail in a different work \cite{Agung epistemic interpretation}.    

\begin{acknowledgments}  
The research is partially funded by John Templeton Foundation (project ID 43297). The opinions expressed in this publication do not necessarily reflect the views of the John Templeton foundation.  I would like to thank Daniel Rohrlich for continuous support and insightful discussion. I also would like to thank the anonymous reviewers for their  constructive comments and suggestions, and Hermawan Dipojono for support and useful discussion.  
\end{acknowledgments}

\end{document}